\documentclass[twocolumn,prl,showpacs,floatfix,preprintnumbers,amsmath,amssymb,superscriptaddress]{revtex4-1}
\usepackage{graphicx}
\usepackage{dcolumn}   
\usepackage{bm}        
\usepackage{amssymb}   
\usepackage{epstopdf}

\begin{document}


\newcommand{\te}{$^{130}$Te}
\newcommand{\tee}{$^{128}$Te}
\newcommand{\nnbb}{$\nu\nu\beta\beta$}
\newcommand{\bbonu}{\ensuremath{0\nu\beta\beta}}
\newcommand{\bboxnu}{\ensuremath{0\nu\chi^0\beta\beta}}
\newcommand{\bbtnu}{\ensuremath{2\nu\beta\beta}}
\newcommand{\qbb}{\ensuremath{Q_{\beta\beta}}}
\newcommand{\mbb}{\ensuremath{\langle m_{\nu} \rangle}}
\newcommand{\bb}{\ensuremath{\beta\beta}}
\newcommand{\MO}{\ensuremath{{}^{100}\rm{Mo}}}
\newcommand{\SE}{\ensuremath{{}^{82}\rm{Se}}}
\newcommand{\ND}{\ensuremath{{}^{150}\rm{Nd}}}
\newcommand{\XE}{\ensuremath{{}^{136}\rm{Xe}}}
\newcommand{\GE}{\ensuremath{{}^{76}\rm{Ge}}}
\newcommand{\AC}{\ensuremath{{}^{228}\rm{Ac}}}
\newcommand{\PO}{\ensuremath{{}^{214}\rm{Po}}}
\newcommand{\TE}{\ensuremath{{}^{130}\rm{Te}}}
\newcommand{\BI}{\ensuremath{{}^{214}\rm{Bi}}}
\newcommand{\BItwo}{\ensuremath{{}^{212}\rm{Bi}}}
\newcommand{\Bi}{\ensuremath{{}^{207}\rm{Bi}}}
\newcommand{\TL}{\ensuremath{{}^{208}\rm{Tl}}}
\newcommand{\BITEN}{\ensuremath{{}^{210}\rm{Bi}}}
\newcommand{\PA}{\ensuremath{{}^{234m}\rm{Pa}}}
\newcommand{\KK}{\ensuremath{{}^{40}\rm{K}}}
\newcommand{\PB}{\ensuremath{{}^{214}\rm{Pb}}}
\newcommand{\CA}{\ensuremath{{}^{48}\rm{Ca}}}
\newcommand{\FT}{\ensuremath{40 \ \textrm{mg/cm}^2}} 
\newcommand{\FTsix}{\ensuremath{60 \ \textrm{mg/cm}^2}} 
\newcommand{\FTtwe}{\ensuremath{20 \ \textrm{mg/cm}^2}} 
\newcommand{\FTeig}{\ensuremath{80 \ \textrm{mg/cm}^2}} 
\newcommand{\Monu}{\ensuremath{\Big|M^{0\nu}\Big|}}
\newcommand{\Mtnu}{\ensuremath{\Big|M^{2\nu}\Big|}}
\newcommand{\Gonu}{\ensuremath{G^{0\nu}(E_0, Z)}}
\newcommand{\Gtnu}{\ensuremath{G^{2\nu}(E_0, Z)}}
\newcommand{\tetaot}{\ensuremath{\theta_{13}}}
\newcommand{\tetatt}{\ensuremath{\theta_{23}}}
\newcommand{\tetaotw}{\ensuremath{\theta_{12}}}
\newcommand{\deltt}{\ensuremath{\Delta_{23}}}
\newcommand{\delot}{\ensuremath{\Delta_{13}}}
\newcommand{\dsun}{\ensuremath{\Delta_{sun}^2}}
\newcommand{\datm}{\ensuremath{\Delta_{atm}^2}}
\newcommand{\Toh}{\ensuremath{T_{1/2}}}
\newcommand{\M}{\ensuremath{M}}
\newcommand{\W}{\ensuremath{W}}
\newcommand{\GA}{\ensuremath{g_A}}
\newcommand{\NA}{\ensuremath{N_A}}
\newcommand{\wbbo}{\ensuremath{N_{\beta\beta}^{0\nu}}}
\newcommand{\wbbt}{\ensuremath{N_{\beta\beta}^{2\nu}}}
\newcommand{\wupper}{\ensuremath{N_{upper}^{2\nu}}}
\newcommand{\epst}{\ensuremath{\epsilon_{\beta\beta}}}
\newcommand{\epsa}{\ensuremath{\epsilon_{\beta\beta}^{accep}}}
\newcommand{\epsr}{\ensuremath{\epsilon_{\beta\beta}^{rec}}}
\newcommand{\epss}{\ensuremath{\epsilon_{\beta\beta}^{sel}}}
\newcommand{\TO}{\ensuremath{T_{1/2}^{0\nu}}}
\newcommand{\TOM}{\ensuremath{T_{1/2}^{0\nu \chi_0}}}
\newcommand{\TT}{\ensuremath{T_{1/2}^{2\nu}}}
\newcommand{\kgy}{\ensuremath{Kg\cdot year}} 
\newcommand{\ev}{\ensuremath{eV}} 
\newcommand{\mev}{\ensuremath{meV}} 
\newcommand{\Mev}{\ensuremath{MeV }} 
\newcommand{\Kev}{\ensuremath{KeV}} 
\newcommand{\mm}{\ensuremath{mm}} 
\newcommand{\cm}{\ensuremath{cm}} 
\newcommand{\kg}{\ensuremath{kg}} 
\newcommand{\ubq}{\ensuremath{\mu Bq / kg}} 
\newcommand{\mbq}{\ensuremath{mBq / kg}} 
\newcommand{\mbqm}{\ensuremath{mBq / m^3}} 
\newcommand{\sn}{SuperNEMO} 
\newcommand{\nt}{NEMO-3} 
\newcommand{\CD}{\ensuremath{^{116}{\rm Cd}}}
\newcommand{\ncl}{90\% C.L.}
\def\NCA{\em Nuovo Cimento}
\def\NIM{\em Nucl. Instrum. Methods}
\def\NIMA{{\em Nucl. Instrum. Methods} \bf A}
\def\NPB{{\em Nucl. Phys.} \bf B}
\def\NPA{{\em Nucl. Phys.} \bf A}
\def\PLB{{\em Phys. Lett.}  \bf B}
\def\PRL{\em Phys. Rev. Lett.}
\def\PRD{{\em Phys. Rev.} \bf D}
\def\ZPC{{\em Z. Phys.} \bf C}
\def\EPJA{{\em Eur. Phys. J.} \bf A}
\def\EPJC{{\em Eur. Phys. J.} \bf C}

\vspace{0.2cm}
\title{Measurement of the Double Beta Decay Half-life of \boldsymbol{$^{130}$}Te~with the NEMO-3 Detector}

\newcommand{\Lebedev}{Nuclear Physics Dept., Lebedev Physical Inst., Leninsky Prospect 53, 117924 Moscow, Russia}
\newcommand{\IHEP}{Inst. for High Energy Physics, Protvino, Moscow Region RU-140284, Russia}
\newcommand{\Texas}{Dept. of Physics, Univ. of Texas, 1 University Station, Austin, TX 78712}
\newcommand{\Wisconsin}{Physics Dept., Univ. of Wisconsin, Madison, WI 53706}
\newcommand{\LAL}{LAL, Universit\'e Paris-Sud 11,  CNRS/IN2P3, Orsay, France}
\newcommand{\IPHC}{IPHC-DRS, Universit\'e Louis Pasteur, CNRS, F-67037 Strasbourg, France}
\newcommand{\INL}{INL, Idaho National Laboratory, Idaho Falls, Idaho 83415, USA}
\newcommand{\ITEP}{ITEP, Institute of Theoretical and Experimental Physics, 117259 Moscow, Russia}
\newcommand{\UdB}{Universit\'e de Bordeaux, 
CENBG, UMR 5797, 
 F-33175 Gradignan, France}
\newcommand{\CNRS}{CNRS/IN2P3,  
CENBG, UMR 5797, 
 F-33175 Gradignan, France}
\newcommand{\JINR}{JINR, Joint Institute for Nuclear Research, 141980 Dubna, Russia}   
\newcommand{\UCL}{University College London, London WC1E 6BT, United Kingdom}
\newcommand{\UMAN}{University of Manchester, Manchester M13 9PL, UK}
\newcommand{\USMBA}{USMBA, Universit\'e Sidi Mohamed Ben Abdellah, 30000 Fes, Morocco}
\newcommand{\UTA}{University of Texas at Austin, Austin, Texas 78712-0264, USA}
\newcommand{\IEAP}{IEAP, Czech Technical University in Prague,  CZ-12800 Prague, Czech Republic}
\newcommand{\LPC}{LPC, ENSICAEN, Universit\'e de Caen, Caen, France}
\newcommand{\IFIC}{IFIC, CSIC - Universidad de Valencia, Valencia, Spain}
\newcommand{\UB}{Universitat Aut\`onoma de Barcelona, Spain}
\newcommand{\Saga}{Saga University, Saga 840-8502, Japan}
\newcommand{\LSCE}{LSCE, CNRS, F-91190 Gif-sur-Yvette, France}
\newcommand{\FMFI}{FMFI, Comenius University, SK-842 48 Bratislava, Slovakia}
\newcommand{\JU}{Jyv\"{a}skyl\"{a} University,  40351 Jyv\"{a}skyl\"{a}, Finland}
\newcommand{\MHC}{MHC, Mount Holyoke College, South Hadley, Massachusetts 01075, USA}
\newcommand{\CU}{Charles University in Prague, Faculty of Mathematics and Physics, CZ-12116 Prague, Czech Republic}
\newcommand{\IC}{Imperial College London, London SW7 2AZ, UK}
\newcommand{\HU}{Hanoi University of Science, Hanoi, Vietnam}
\newcommand{\dec}{Deceased}

\affiliation{\IPHC}
\affiliation{\LAL}
\affiliation{\INL}
\affiliation{\ITEP}
\affiliation{\UCL}
\affiliation{\UdB}
\affiliation{\CNRS}
\affiliation{\JINR}
\affiliation{\LPC}
\affiliation{\UMAN}
\affiliation{\USMBA}
\affiliation{\UTA}
\affiliation{\IEAP}
\affiliation{\IFIC}
\affiliation{\HU}
\affiliation{\Saga}
\affiliation{\LSCE}
\affiliation{\IC}
\affiliation{\FMFI}
\affiliation{\JU}
\affiliation{\MHC}
\affiliation{\CU}

\author{R.~Arnold}
\affiliation{\IPHC}
\author{C.~Augier}
\affiliation{\LAL}
\author{J.~Baker}
\affiliation{\INL}
\affiliation{\dec}
\author{A.S.~Barabash}
\affiliation{\ITEP}
\author{A.~Basharina-Freshville}\affiliation{\UCL}
\author{S.Blondel}\affiliation{\LAL}
\author{M.~Bongrand}
\affiliation{\LAL}
\author{G.~Broudin-Bay}
\affiliation{\UdB}\affiliation{\CNRS}
\author{V.~Brudanin}
\affiliation{\JINR}
\author{A.J.~Caffrey}
\affiliation{\INL}
\author{A.~Chapon}  
\affiliation{\LPC}
\author{E.~Chauveau}
\affiliation{\UMAN}
\author{D.~Durand}  
\affiliation{\LPC}
\author{V.~Egorov}
\affiliation{\JINR}
\author{R.~Flack}
\affiliation{\UCL}
\author{X.~Garrido}
\affiliation{\LAL}
\author{J.~Grozier}\affiliation{\UCL}
\author{B.~Guillon}  
\affiliation{\LPC}
\author{Ph.~Hubert}
\affiliation{\UdB}\affiliation{\CNRS}
\author{C.M.~Jackson}  
\affiliation{\UMAN}
\author{S.~Jullian}
\affiliation{\LAL}
\author{M.~Kauer}
\affiliation{\UCL}
\author{A.~Klimenko}
\affiliation{\JINR}
\author{O.~Kochetov}
\affiliation{\JINR}
\author{S.I.~Konovalov}
\affiliation{\ITEP}
\author{V.~Kovalenko}
\affiliation{\UdB}\affiliation{\CNRS}\affiliation{\JINR}
\author{D.~Lalanne}
\affiliation{\LAL}
\author{T.~Lamhamdi}
\affiliation{\USMBA}
\author{K.~Lang}
\affiliation{\UTA}
\author{Z.~Liptak}
\affiliation{\UTA}
\author{G.~Lutter}
\affiliation{\UdB}\affiliation{\CNRS}
\author{F.~Mamedov}
\affiliation{\IEAP}
\author{Ch.~Marquet}
\affiliation{\UdB}\affiliation{\CNRS}
\author{J.~Martin-Albo}
\affiliation{\IFIC}
\author{F.~Mauger}
\affiliation{\LPC}
\author{J.~Mott}\affiliation{\UCL}
\author{A.~Nachab}
\affiliation{\UdB}\affiliation{\CNRS}
\author{I.~Nemchenok}
\affiliation{\JINR}
\author{C.H.~Nguyen}  
\affiliation{\UdB}\affiliation{\CNRS}\affiliation{\HU}
\author{F.~Nova}
\affiliation{\UTA}
\author{P.~Novella}
\affiliation{\IFIC}
\author{H.~Ohsumi}
\affiliation{\Saga}
\author{R.B.~Pahlka}
\affiliation{\UTA}
\author{F.~Perrot}
\affiliation{\UdB}\affiliation{\CNRS}
\author{F.~Piquemal}
\affiliation{\UdB}\affiliation{\CNRS}
\author{J.L.~Reyss}
\affiliation{\LSCE}
\author{B.~Richards}\affiliation{\UCL}
\author{J.S.~Ricol}
\affiliation{\UdB}\affiliation{\CNRS}
\author{R.~Saakyan}
\affiliation{\UCL}
\author{X.~Sarazin}
\affiliation{\LAL}
\author{L.~Simard}
\affiliation{\LAL}
\author{F.~\v{S}imkovic}
\affiliation{\FMFI}
\author{Yu.~Shitov}
\affiliation{\JINR}\affiliation{\IC}
\author{A.~Smolnikov}
\affiliation{\JINR}
\author{S.~S\"{o}ldner-Rembold}
\affiliation{\UMAN}
\author{I.~\v{S}tekl}
\affiliation{\IEAP}
\author{J.~Suhonen}
\affiliation{\JU}
\author{C.S.~Sutton}
\affiliation{\MHC}
\author{G.~Szklarz}
\affiliation{\LAL}
\author{J.~Thomas}
\affiliation{\UCL}
\author{V.~Timkin}
\affiliation{\JINR}
\author{S.~Torre}\affiliation{\UCL}
\author{V.I.~Tretyak}
\affiliation{\IPHC}\affiliation{\JINR}
\author{V.~Umatov}
\affiliation{\ITEP}
\author{L.~V\'{a}la}
\affiliation{\IEAP}
\author{I.~Vanyushin}
\affiliation{\ITEP}
\author{V.~Vasiliev}
\affiliation{\UCL}
\author{V.~Vorobel}
\affiliation{\CU}
\author{Ts.~Vylov}
\affiliation{\JINR}
\affiliation{\dec}
\author{A.~Zukauskas}
\affiliation{\CU}

\collaboration{The NEMO-3 Collaboration}
\noaffiliation

\date{\today}

\begin{abstract}
We report results from the NEMO-3 experiment
based on an exposure of 1275 days with  $661$~g of $^{130}$Te in the form
of enriched and natural tellurium foils. The double beta 
decay rate of $^{130}$Te is found to be greater than zero with a significance of $7.7$ 
standard deviations and the half-life is measured to be
$
\TT = (7.0 \pm 0.9(\rm{stat})\; \pm 1.1(\rm{syst}) )\; \times 10^{20}\; \rm yr.
$
This represents the most precise measurement of this half-life yet published 
and the first real-time observation of this decay. 
\end{abstract}
\pacs{23.40.-s, 21.10.Tg, 14.60.Pq}
\maketitle

The first evidence of double beta decay (\bbtnu) appeared in $1950$ 
through the observation of  $^{130}$Xe~from the decay of \TE~in rock samples~\cite{Inghram}.
This result was met with scepticism for the ensuing $15$ years until the results of 
a number of other geochemical experiments 
began to confirm the observation. There was, however, significant disagreement 
between two distinct sets of these measurements that was not immediately resolved.  
Several groups measured a long half-life of $\approx 2.7\times 10^{21}$~yr~\cite{Kirsten,Bernatowicz} 
while others obtained a significantly shorter half-life of 
$\approx0.8\times10^{21}$~yr~\cite{Manuel,Lin,Takaoka,Takaoka2}. 
One hypothesis to explain the difference is based on the observation that 
shorter 
half-lives
were measured in rock of relatively young age ($\sim 10^{7}-10^{8}$~yr), while the longer 
half-lives
were measured in relatively old rock ($\sim 10^{9}$~yr)~\cite{Kirsten,Manuel}. 
It has even been suggested that there is a time dependence 
in the value of the weak interaction coupling constant~\cite{sacha}.
Recent papers~\cite{Meshik} attempt to explain the long-held discrepancy between these 
measurements as being caused by catastrophic xenon loss in the older samples. 
To date, the only direct evidence for the \TE~ \bbtnu~ process in an 
experiment comes from MIBETA, which reported a half-life of
$(6.1\pm 1.4(\rm{stat}) ^{+2.9}_{-3.5}(\rm{syst})) \times 10^{20}$ yr~\cite{mibeta} 
by comparing different crystals isotopically enriched in \TE~and \tee, 
assuming that any difference in rate was due to \bbtnu\ events 
(\tee~has a much longer $\bbtnu$~half-life).
However, a systematic uncertainty of about $50\%$ 
rendered this result somewhat inconclusive.
In this Letter, we present the first direct, high-precision 
measurement of \bbtnu\ decay of  \TE\  with the \nt\  
detector.
In addition, a search for neutrinoless double beta decay 
(\bbonu) and for the decay with Majoron emission (\bboxnu) is reported.

The \nt\ detector is located in the Modane Underground Laboratory. 
The detector~\cite{TDR} contains almost 9~kg of seven different $\beta\beta$ isotopes 
in the form of thin foils. It provides direct detection of electrons 
from the double beta decay by the use of a tracking device 
based on open Geiger drift cells and a calorimeter made of plastic 
scintillator blocks coupled to low-radioactive photomultipliers (PMTs). 
For $1$~MeV electrons the timing resolution is 250~ps and the energy 
resolution (full width at half maximum) is about $15\%$. 
A magnetic field surrounding the detector 
provides identification of electrons by the curvature of their tracks. 
In addition to the electron and photon identification through tracking and calorimetry, 
the calorimeter measures the energy and the arrival time of these 
particles while the tracking chamber can measure the time of delayed 
tracks associated with the initial event for up to 700~$\mu$s. 
The calorimeter energy scale is calibrated approximately once per month using 
a \Bi~ source providing conversion electrons of 482~keV and 976~keV (K-lines). 
Stability of the calorimeter response is surveyed twice a day by a laser system. 
The data presented in this Letter correspond to 1275 days of data taking
between October 2004 and December 2009. 
Two different foils are used in the analysis: a Te
foil, enriched at a level of $89.4\pm0.5\%$ corresponding
to $454$~g of \te, 
and a natural Te foil which contains $33.8\%$  \te, corresponding to
$207$~g of \te.

When searching for rare processes, the background estimation is paramount as 
it will limit the final sensitivity of the experiment. An exhaustive 
program of work has been carried out to measure the very large number 
of sources of background present in the \nt\ detector.
The method of the background measurement and its validation with 
a highly radiopure Cu foil is described in~\cite{BGRPAPER}.  
There are three categories of background:
the external background, originating from radioactivity outside  
the tracking chamber; the tracking volume background, which includes radon in the 
tracking gas and the drift cell wire contamination; and the internal 
background due to radioactive impurities inside the source foils whose dominant 
isotopes are $^{40}$K, \PA, $^{210}$Bi, $^{214}$Bi and $^{208}$Tl.
The different background contributions are estimated by measuring independent 
event topologies, both for enriched and natural Te. 

\begin{figure}[ht]
\begin{center}
\includegraphics[width=0.22\textwidth]{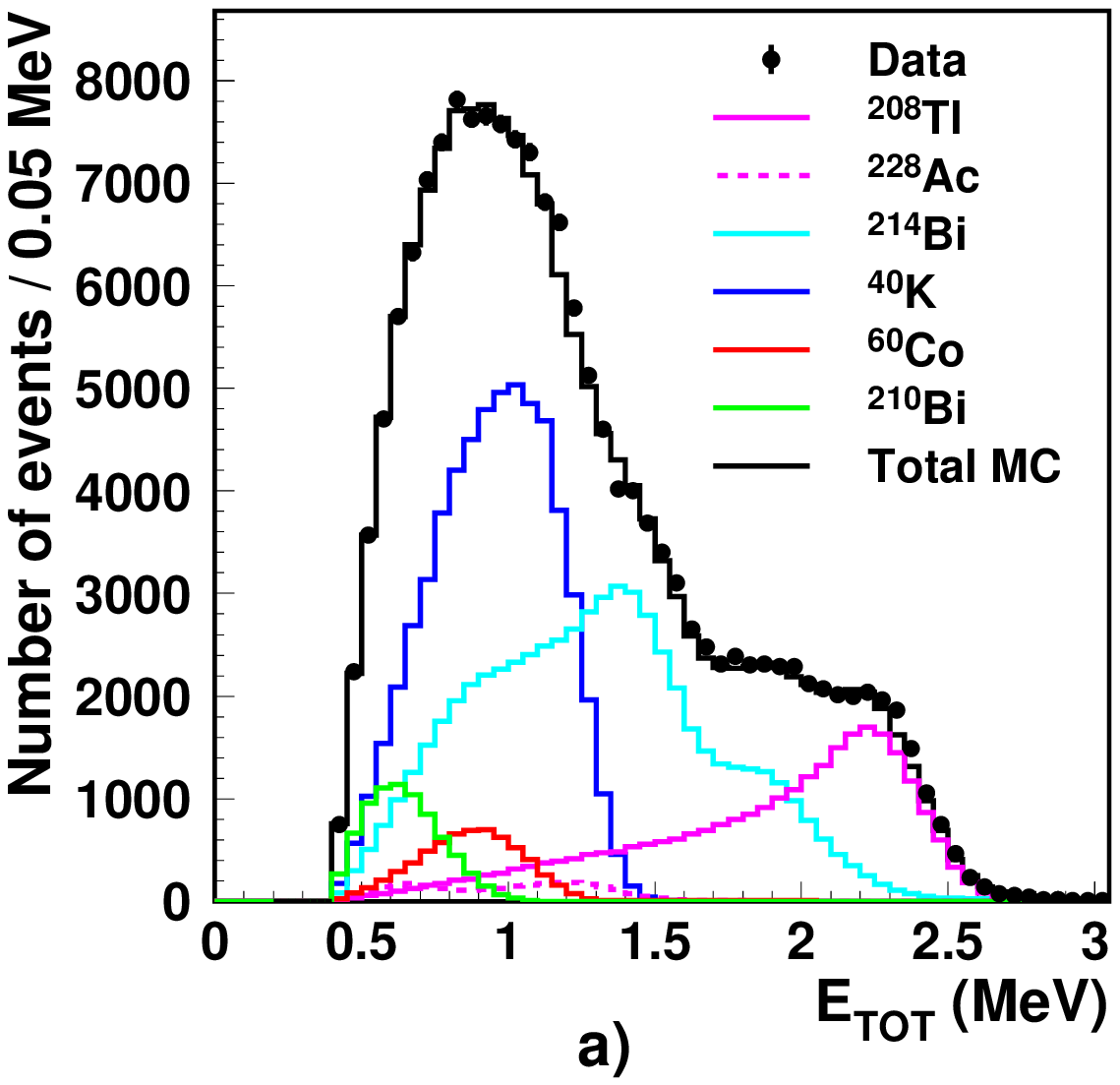}
\includegraphics[width=0.22\textwidth]{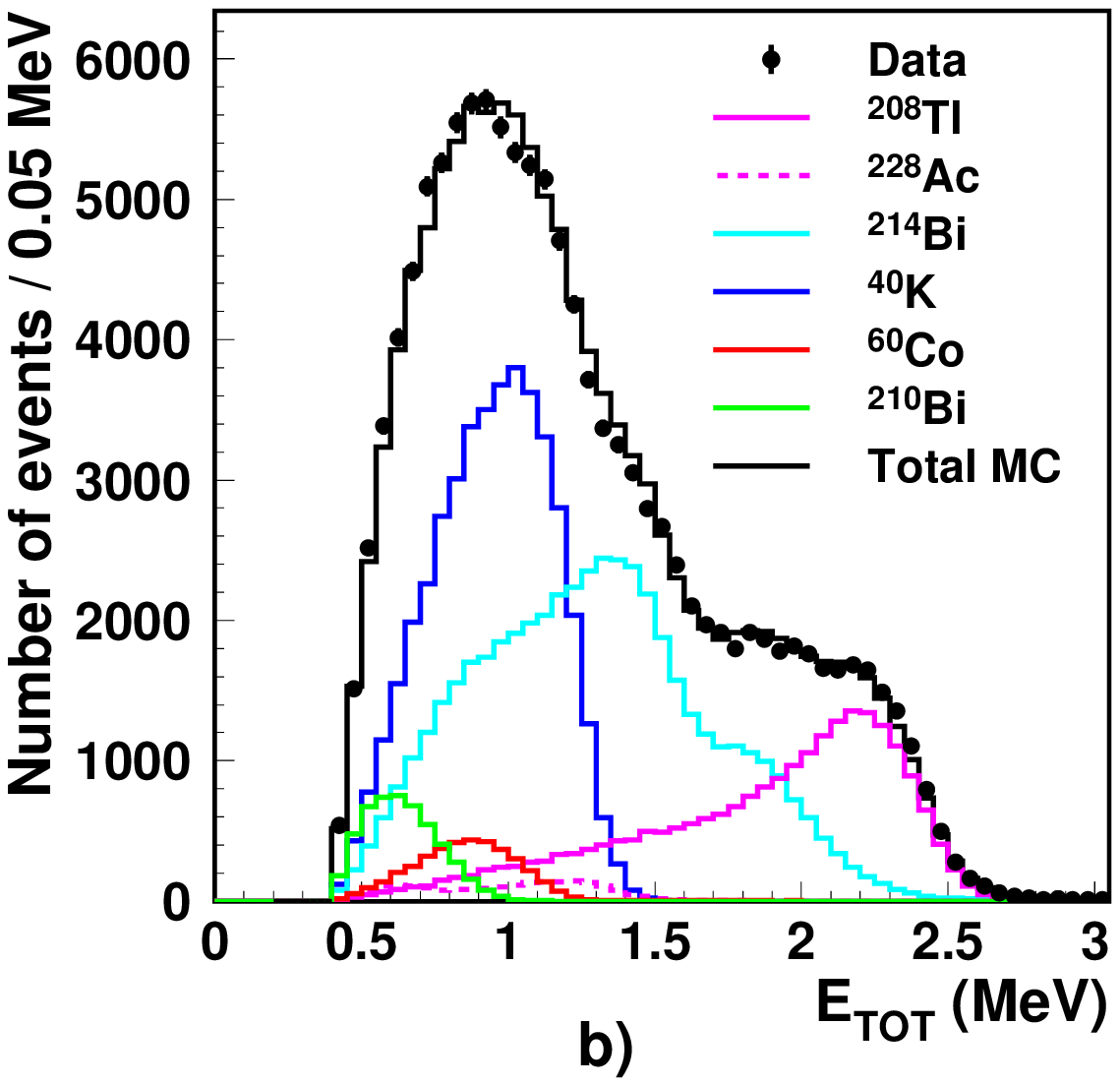}
\includegraphics[width=0.22\textwidth]{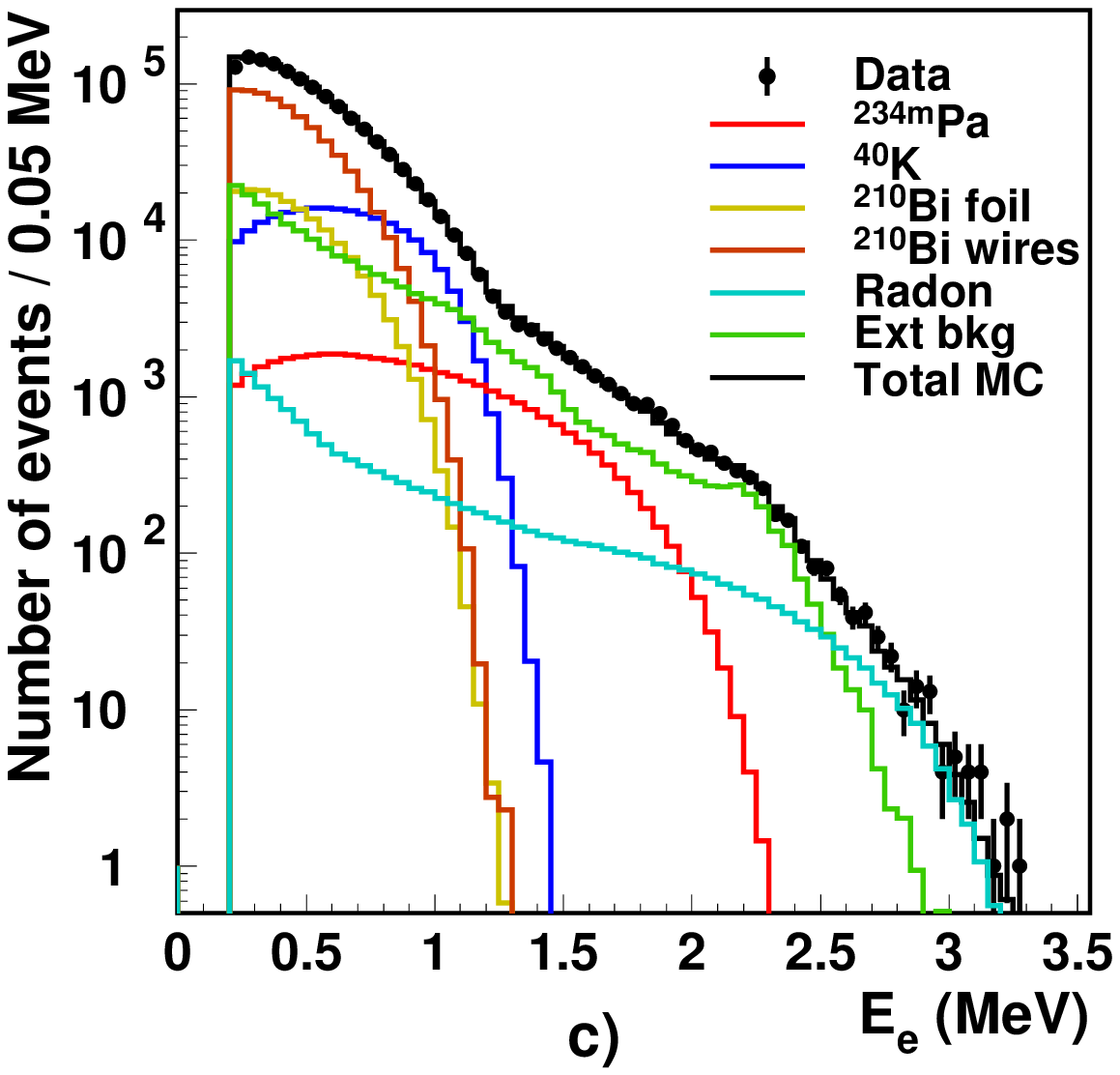}
\includegraphics[width=0.22\textwidth]{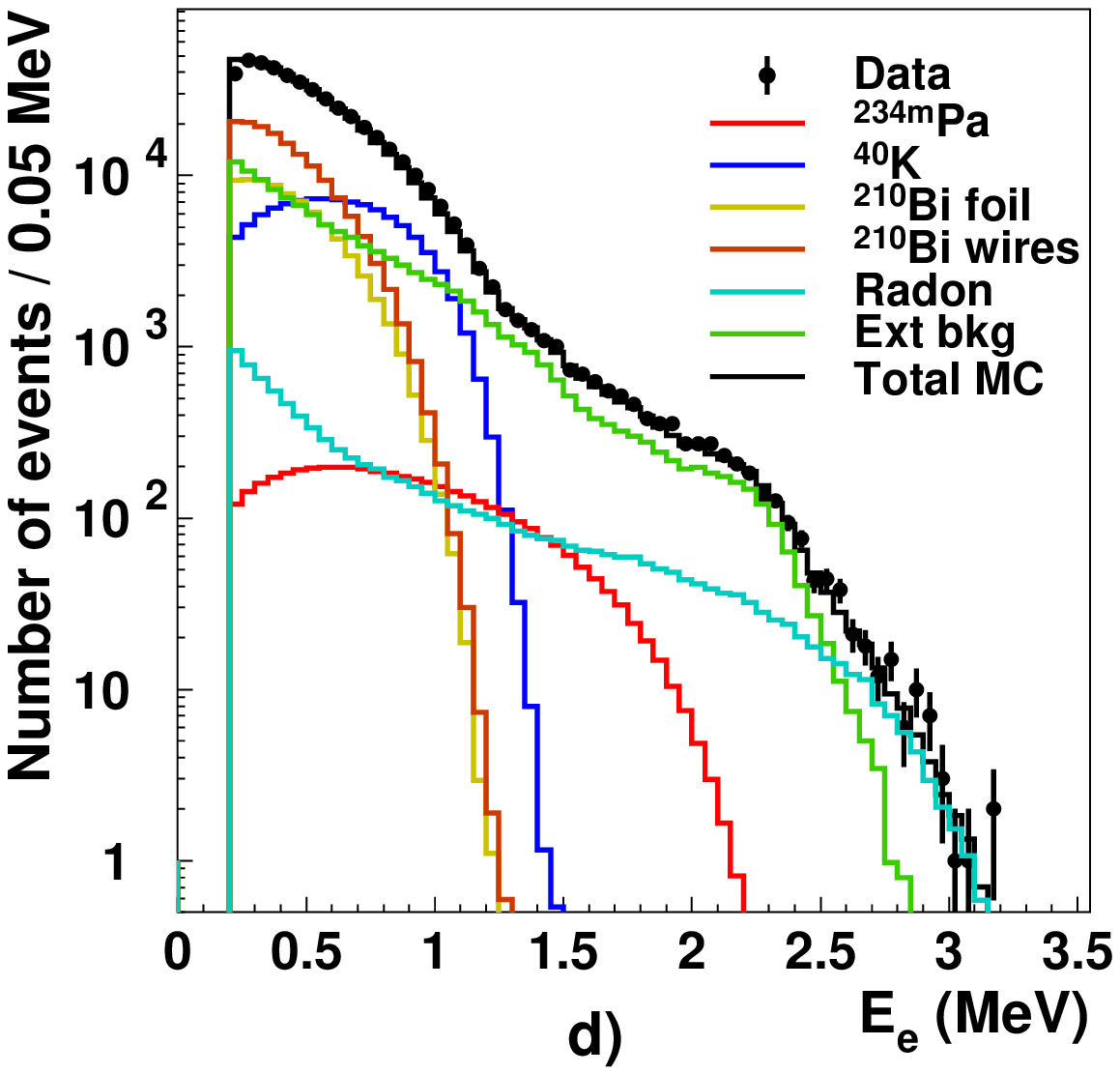}
\parbox[h]{8cm}{
\caption{(Color online) Energy sum distribution for crossing-electron events 
for (a) $^{130}$Te and (b) $^{\rm nat}$Te (b); Energy distribution 
of electron events coming from the source foil for (c) $^{130}$Te 
and (d) $^{\rm nat}$Te. 
Dots correspond to the data and histograms to the 
fit of the background model.
\label{fig:bkg}}
}
\end{center}
\end{figure}

The external background originates from components outside the tracking volume. 
PMTs are the main contributors since they have glass and 
electronic components that, though at a very low level, contain traces 
of \BI, \TL\ and \KK. 
The external background is measured with $\gamma$-ray Compton scattering 
in the scintillators, 
either producing an electron that crosses the tracking chamber or depositing 
energy in one scintillator, followed by an electron emitted from the source 
to another scintillator. 
The reliability of the external background model is illustrated by the 
energy distributions of the one-electron crossing events in Figs.~\ref{fig:bkg}(a) and (b).

The source foil activities in \PA, $^{40}$K and $^{210}$Bi are determined with 
single-electron events coming from the foil.
The energy distribution of the observed events and the result of the fit of the 
different components of the background  
are presented 
in Figs.~\ref{fig:bkg}(c) and (d). The good agreement between the data and the fit 
demonstrates the reliability of the internal background model.
The foil activity in $^{214}$Bi is measured 
using events with a single electron accompanied by a delayed $\alpha$-particle 
track. This topology is a signature of the $\beta$ decay of \BI\ to \PO\
followed by $\alpha$ decay of \PO\ to $^{210}$Pb.
The foil activity in $^{208}$Tl is measured with events that contain 
one electron and either two or three photons emitted from the foil.
The results of measurements of the internal contamination by \BI\ and \TL\  are reported 
in~\cite{BGRPAPER}. 
The measured foil activities are summarized in Table~\ref{tab:bi_tl_act}.
\begin{tiny}
\begin{table}[ht]
	\begin{center}
	\begin{tabular}{|l|c|c|} \hline
	 Impurity & \TE\ &  $^{\rm nat}$Te\\  
	\hline
	\TL\ &$0.17 \pm 0.04$    &$0.24 \pm 0.04$  \\
	\BI\ & $0.29 \pm 0.05$  & $0.25 \pm 0.12$ \\
	\PA\ &$2.49 \pm 0.05$   &$0.73 \pm 0.04$ \\
	\BITEN\ &$19.9 \pm 0.4$ & $18.4 \pm 0.3$ \\
	 \KK\ &$14.7 \pm 0.2$    &$12.3 \pm 0.2$  \\
 	\hline 
	\end{tabular}
	\caption{Background contaminations measured in the Te source foils (in mBq).
	\label{tab:bi_tl_act}}
	\end{center}
\end{table}
\end{tiny}

The background from single $\beta$ decay in the tracking 
volume is of importance if the decay occurs near the foil.
The main source of this background is 
due to daughters of radon: $^{214}$Pb, \BI\ and $^{210}$Bi.
The radon activity in the tracking chamber is measured using 
$e\alpha$ events as for the measurement of internal $^{214}$Bi background.
The distribution of these events is 
measured as a function of the location in the tracking volume.
For the data presented here, the mean 
$^{222}$Rn activity in the whole gas volume is $209 \pm 2$~mBq.
Unlike the preceding radon daughters, $^{210}$Pb has a long half-life of
about 22 years. It is therefore not in equilibrium 
with $^{222}$Rn in the tracking volume and most $^{210}$Pb 
was deposited during the construction of the detector.
$^{210}$Bi produced in $\beta$ decay of 
$^{210}$Pb contributes to the low-energy background below 1~MeV. 
The $^{210}$Pb deposition on drift cell wires, measured by
detecting electrons from $^{210}$Bi $\beta$ decay, was found
to be vary significantly in different sectors~\cite{BGRPAPER}.
In contrast to $^{222}$Rn
the concentration of $^{220}$Rn in \nt\ is very small and its contribution 
to the total background in the \TE\ sectors 
is less than $1\%$.

The measured activities are used to estimate the background contribution 
in the two-electron channel with a Monte Carlo (MC) simulation.
Signal and background MC events are generated using 
a {\sc geant}-based simulation~\cite{Geant} 
of the detector with the initial kinematics given by the event 
generator {\sc decay0}~\cite{Decay0}.

The two-electron events are selected with the following requirements.
Two tracks of a length greater that 50 cm with curvature corresponding to a
negative charge are reconstructed. 
Both tracks originate from a common vertex in the foil and terminate 
in isolated scintillators with a single energy deposit greater than 0.2~MeV.  
The time-of-flight information is consistent with the hypothesis that two 
electrons were emitted from the same point on the source foil. 
No photon or delayed $\alpha$ track is detected 
in the event.
These selection criteria lead to a \bbtnu\  detection efficiency of $3.5\%$ for enriched  
\TE\ and $2.8\%$ for $^{\rm nat}$Te, where the difference is mainly due to the source 
foil thickness. 
Several factors contribute to this low efficiency, but the most important are 
the geometrical acceptance of the detector, the effect of the energy threshold 
and the tracking algorithm inefficiencies.

\begin{figure}[t]
\begin{center}
\includegraphics[width=0.22\textwidth]{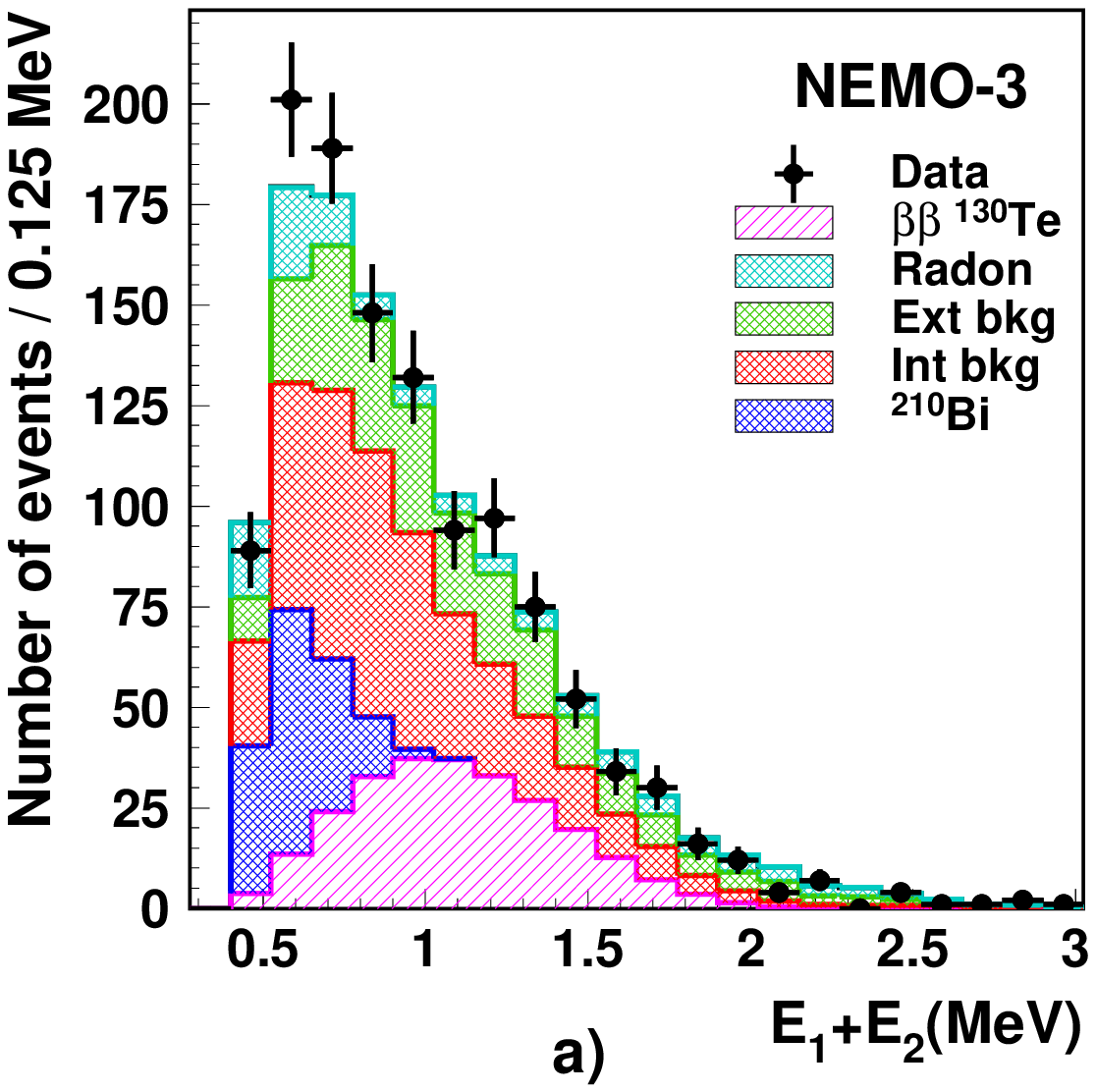}
\includegraphics[width=0.22\textwidth]{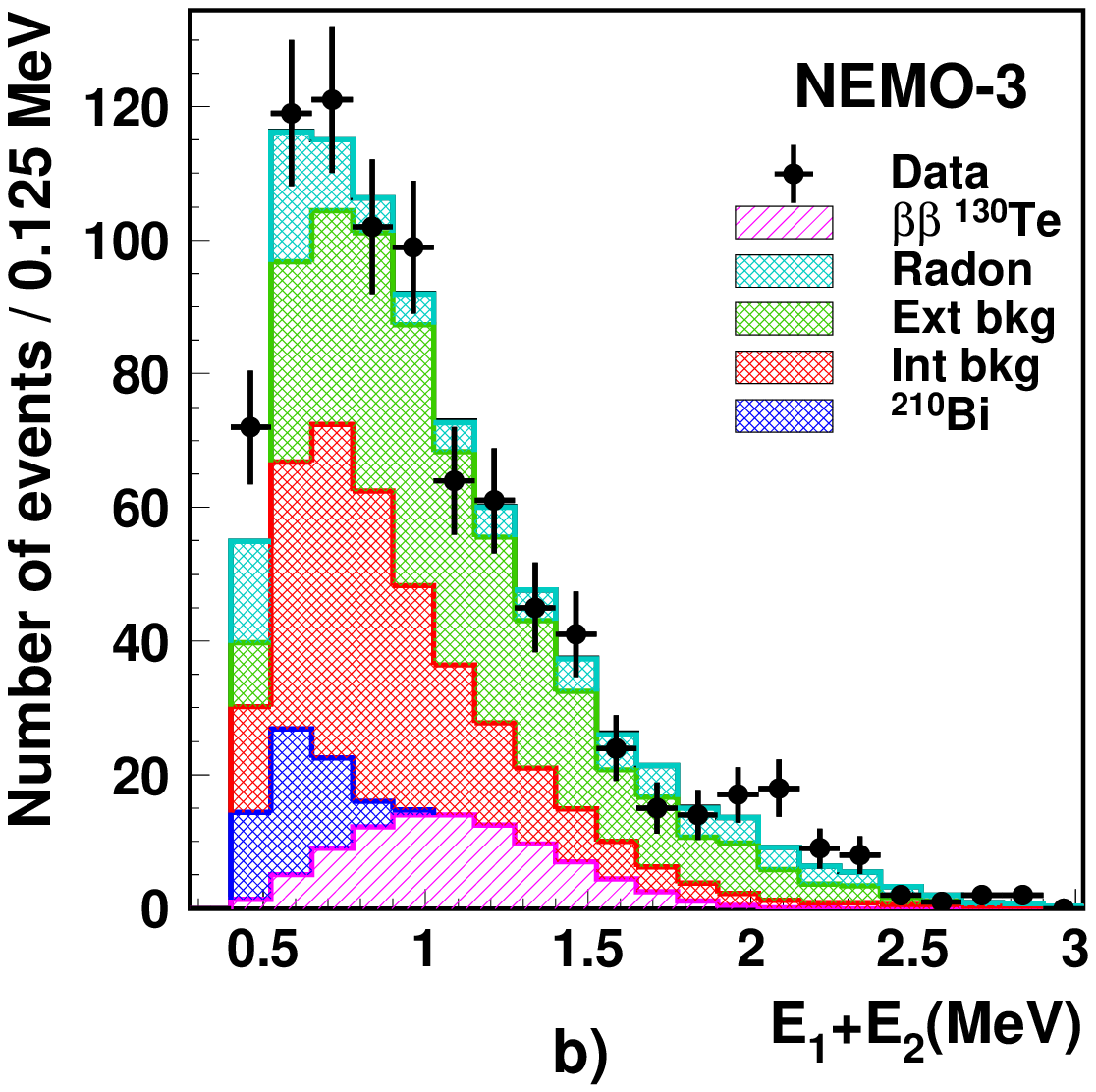}
\includegraphics[width=0.22\textwidth]{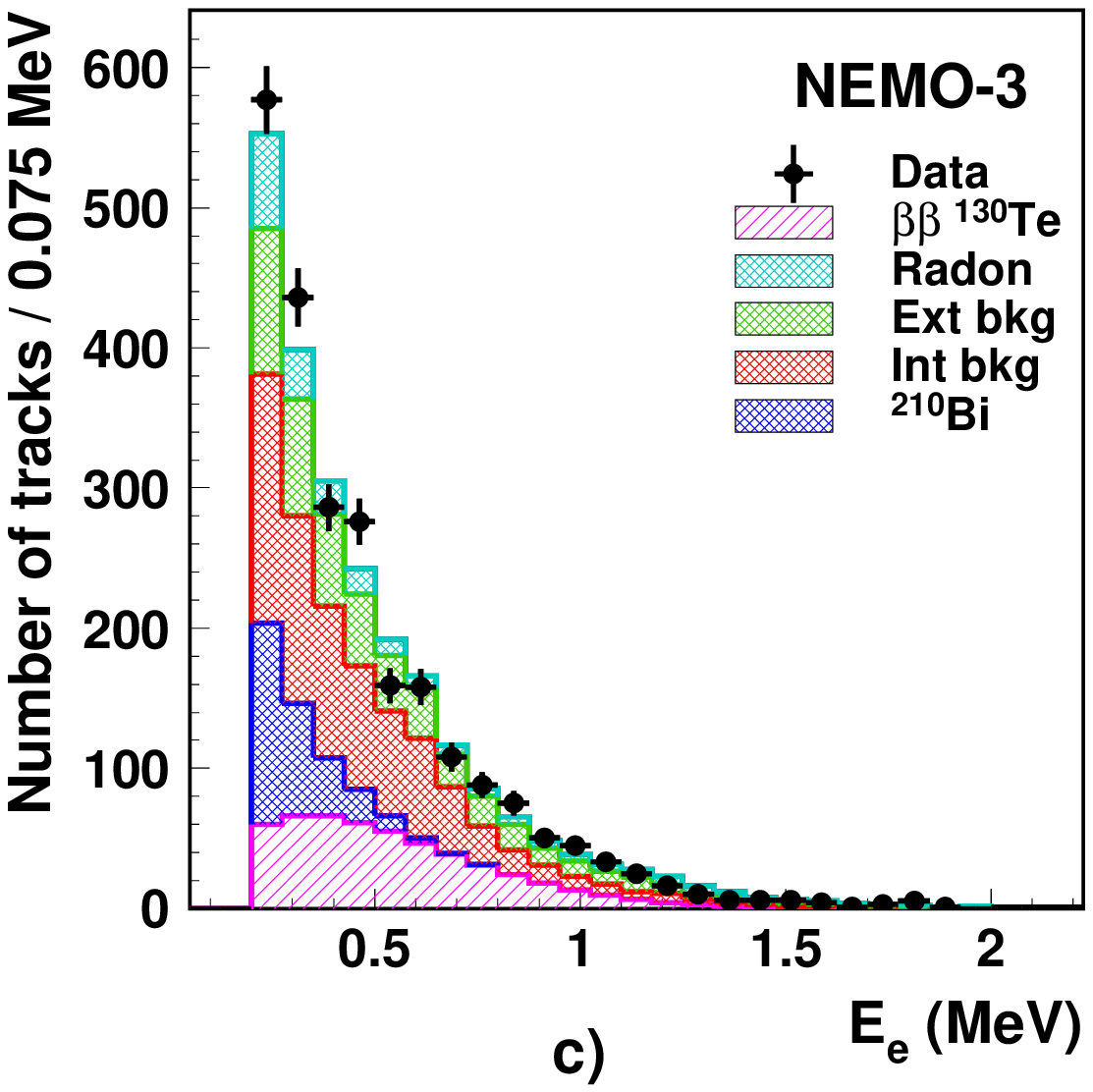}
\includegraphics[width=0.22\textwidth]{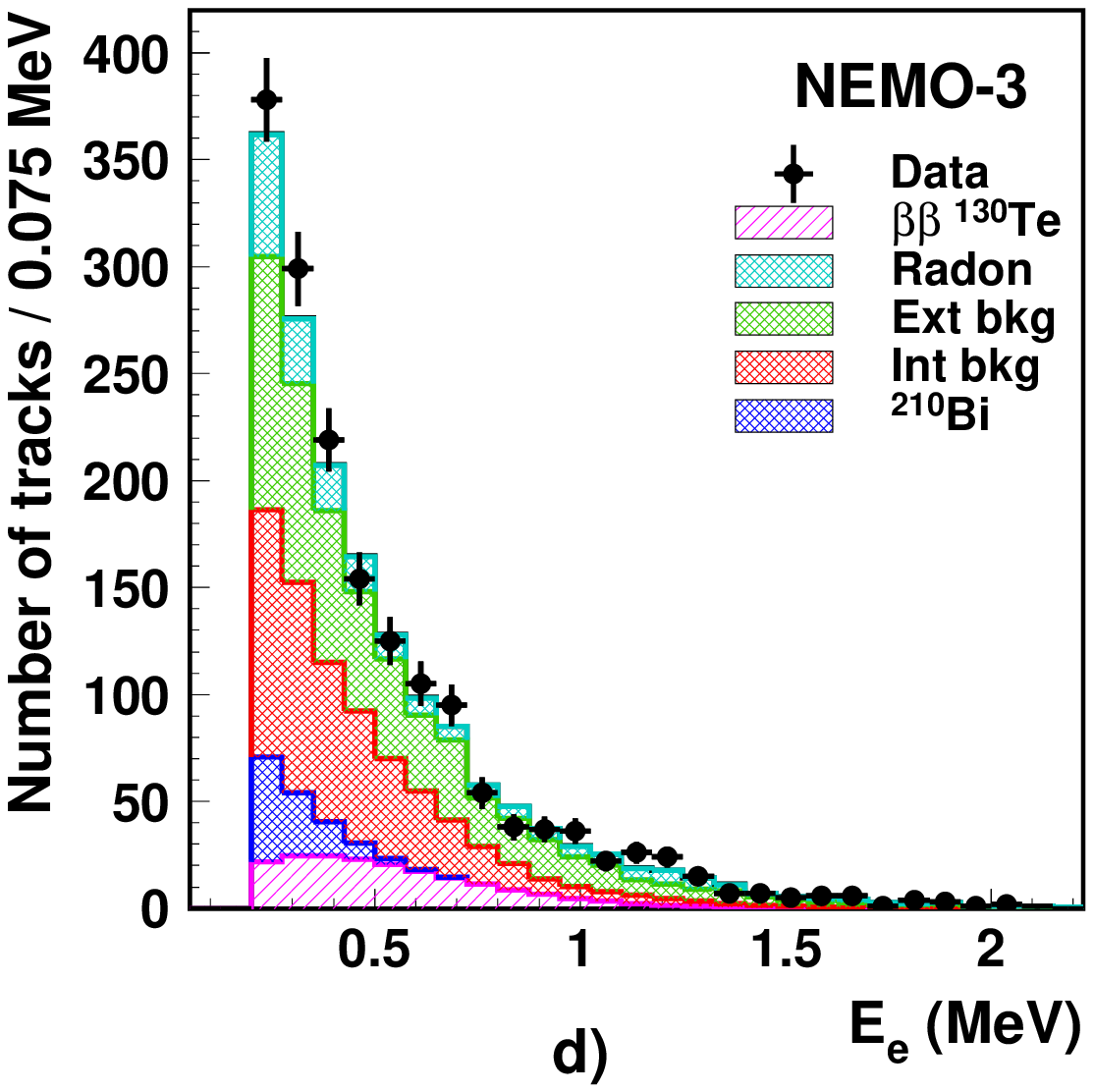}
\includegraphics[width=0.22\textwidth]{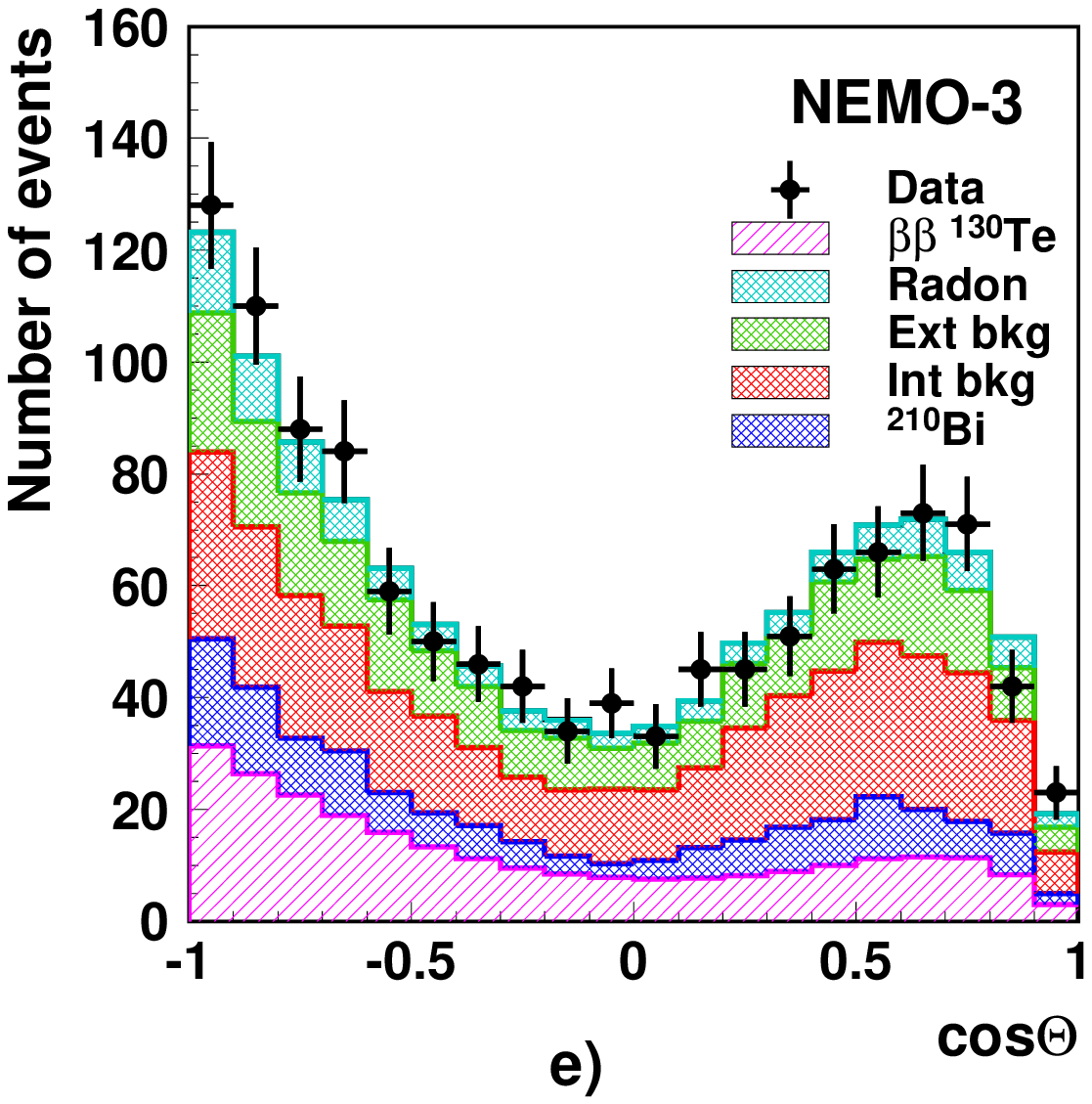}
\includegraphics[width=0.22\textwidth]{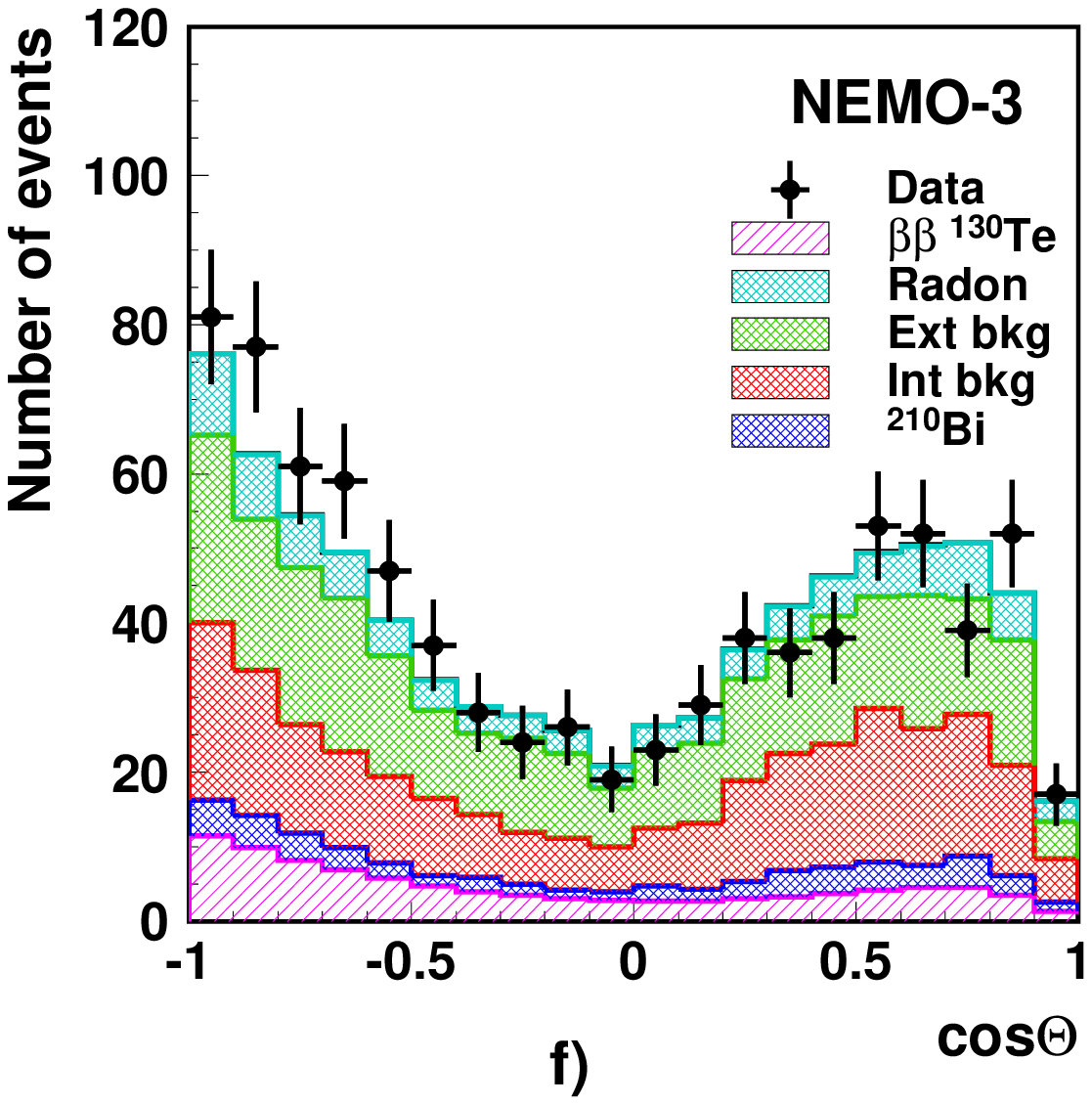}
\parbox[h]{8cm}{
\caption{(Color online)
(a,b)  Distribution of the sum of the electron energies; 
(c,d) individual electron energy; and 
(e,f) cosine of the angle between the two electron tracks for
two-electron events selected from the two Te~foils:
(a,c,e) enriched in \TE\ and (b,d,f) natural Te.
\label{fig:ee_te_130}}
}
\end{center}
\end{figure}

We measure the \TE\ half-life with data from the enriched Te source foil
by performing a likelihood fit to the binned energy 
sum distribution in the interval $[0.9-2]$~MeV.
This interval is chosen using the MC simulation to maximize the signal 
significance. It reduces the \bbtnu\ efficiency by a factor of 0.7.
The result of the fit is presented in Figs.~\ref{fig:ee_te_130}(a), (c), and (e).
Using the MC simulation the number of background events
in the interval $[0.9-2]$~MeV is estimated to be
 $363 \pm 25$ events, of which
$141$ events are associated with the external background, $179$ with
the internal  background and $43$ with radon induced background.
The number of events in excess of the background in the interval $[0.9-2]$~MeV is
determined to be
\begin{equation}
n(\bbtnu )= 178 \pm 23,
\end{equation}
with a signal significance of $7.7$ standard deviations and a signal-to-background 
ratio  of $S/B=0.5$. This corresponds to a \TE\ half-life of
\begin{equation}
\TT = (7.0 \pm 0.9 )\;\times  10^{20}\; \rm yr.
\end{equation}

The main systematic uncertainty on the measured \TE\ half-life 
is associated with the background estimation and is due to the small 
signal-to-background ratio. The uncertainty on the number of expected 
background events has been obtained by applying the largest variations 
of the component activities in the background model.
The corresponding uncertainty on the \TE\ half-life is $14\%$. 
Another systematic uncertainty is associated with the two-electron 
detection efficiency in \nt\ which is found to be 
correct within an accuracy of $5\%$. This uncertainty is determined
with a calibrated $^{207}$Bi source and a 
dedicated $^{90}$Sr source which decays to $^{90}$Y, a pure $\beta$ emitter 
of $Q_{\beta}=2.28$~MeV. 
Finally, the source foil thickness and the 
{\sc geant} model of electron energy 
losses in dense thin media contribute a systematic uncertainty of $4\%$
which is estimated by comparing signals from
metallic and composite \MO\ source foils 
\cite{excited}. 
The total systematic uncertainty of $15\%$ is obtained by adding 
the individual contributions in quadrature.

The measurement of the $\beta\beta2\nu$ half-life is verified 
using the natural Te foil.  
The energy and angular distributions of the two-electron events of the 
natural Te data are presented in Fig.~\ref{fig:ee_te_130}b,d,f and are 
compared  with the expected MC distribution using the measured half-life 
$\TT = 7.0 \times  10^{20}$~yr and the background model of the natural Te foil. 
There are $65 \pm 8$ $\beta\beta$ events 
and $316 \pm 28$ background events expected in the electron energy 
sum interval $[0.9-2]$~MeV. 
The total number of $381 \pm 29$ expected events is in good agreement with 
the 377 observed events.

The \te\ data (Fig~\ref{fig:ee_te_130}a) are also used to set a limit on the \bbonu\ 
and \bboxnu\ processes with the CL$_s$ method~\cite{Junk}. 
The method uses the full information of the binned energy sum distribution 
for signal and background, as well as the statistical and systematic uncertainties and their correlations 
as described in ~\cite{Nd-150}. 

The total efficiency to detect \bbonu\  decay of \te\ 
is estimated to be $(13.9\pm0.7)\%$ yielding a limit of 
\begin{equation}
\TO > 1.3 \times 10^{23}\; \rm yr \; (90\%~C.L.),
\end{equation}
which is an order of magnitude less stringent than the limit obtained by the CUORICINO 
Experiment~\cite{Cuoricino} based on 11~kg of \TE.

The detection efficiency for the decay with ordinary (spectral index $n=1$) 
Majoron emission 
(see discussion in~\cite{Majoron} and references therein) is $(9.6\pm0.5)\%$ 
and the limit is determined to be
\begin{equation}
\TOM > 1.6 \times 10^{22}\; \rm yr \; (90\%~C.L.),
\end{equation}
which is a factor of $7$ more stringent than the previous best limit from 
MIBETA~\cite{mibeta}. 
The corresponding limit on the coupling constant of the Majoron to the neutrino 
is $\rm{g_{ee}}<(0.6-1.6)\times 10^{-4}$ (using 
nuclear matrix elements 
from~\cite{simkovic,kortelainen,Caurie,Barea,Chaturvedi})
and is comparable with the best present limits. 

In summary, the \bbtnu\ decay \TE\ half-life measured with the \nt\ detector is
\begin{equation}
\TT = (7.0 \pm 0.9\; (\rm{stat})\; \pm 1.1\;(\rm{syst}) )\;\times 10^{20}\; \rm yr.
\end{equation}
With this result, 
the corresponding nuclear matrix element can be extracted according to 
\begin{equation}
(\TT)^{-1} = \rm{G}^{2\nu}|M^{2\nu}|^{2},
\end{equation} 
where $G^{2\nu} = 4.8 \times 10^{-18}\; \rm{yr}^{-1}$ (for $g_{A}=1.254$) 
is the known phase space
factor~\cite{psf}, which yields the result 
$M^{2\nu}=0.017\pm 0.002$ (scaled by the electron mass). 
This value for $M^{2\nu}$ may be used to fix the g$_{pp}$
parameter of the QRPA theory, which corresponds to the strength of the nucleon-nucleon 
interaction inside the nucleus. 
It has been suggested that this will improve the M$^{0\nu}$ 
calculations~\cite{rodin,simkovic,kortelainen}.

The  \nt\  result for the \TE\ half-life is consistent with the 
geological measurements made in younger rock samples and is the most 
precise measurement of this isotope half-life to date.  

We thank the staff at the Modane Underground Laboratory 
for its technical assistance in running the experiment and Vladimir
Tretyak for providing the Monte Carlo event generator~\cite{Decay0}.
We acknowledge support by the Grants Agencies 
of the Czech Republic, RFBR (Russia), STFC (UK) and NSF (USA).


\end{document}